\documentclass[twocolumn,prd,nofootinbib,aps,prl,floats,floatfix,amsmath,amssymb,longbibliography,secnumarab
ic]{revtex4-1} %
\usepackage[final]{graphicx}
\usepackage{hyperref}
\usepackage{amsmath}
\usepackage{bbm}
\usepackage{amsfonts}
\usepackage{amssymb}
\usepackage{latexsym}
\usepackage{graphicx}
\usepackage[english]{babel}
\usepackage{multirow}
\usepackage{float}
\usepackage{url}
\usepackage{slashed}
\usepackage{xcolor} 
\usepackage[utf8]{inputenc}
\usepackage[nice]{nicefrac}

\def\sfrac#1#2{{\textstyle{#1\over #2}}}
\newcommand{\be}{\begin{equation}}
\newcommand{\ee}{\end{equation}}
\newcommand{\ba}{\begin{array}}
\newcommand{\ea}{\end{array}}
\newcommand{\bea}{\begin{eqnarray}}
\newcommand{\eea}{\end{eqnarray}}
\newcommand{\sss}{\scriptscriptstyle}

\newcommand{\nn}{\nonumber}

\begin{document}

\title{Viable secret neutrino interactions with ultralight dark matter}

\author{James M.\ Cline}
\affiliation{McGill University, Department of Physics, 3600 University St.,
Montr\'eal, QC H3A2T8 Canada}
\begin{abstract}
Several anomalies in neutrino oscillation experiments point to the
existence of a $\sim 1\,$eV sterile neutrino $\nu_s$ mixing with 
$\nu_e$ at the level of $U_{e4}\cong 0.1$, but such a neutrino is
strongly disfavored by constraints on additional light
degrees of freedom ($\delta N_{\rm eff}$) and total neutrino
mass ($\sum_\nu m_\nu$) from cosmology.
``Secret neutrino interactions'' that have been invoked to suppress
the cosmological production of $\nu_s$ typically falter, but recently
it was pointed out that $\nu_s$ could get a large mass in the early
universe by coupling to ultralight dark matter $\phi$, which can robustly 
suppress its production.  The model has essentially two free 
parameters: $m_\phi$, and $m_{s,0}$, the mass of the sterile neutrino
at early times, enhanced by its coupling to $\phi$.  I determine
the parameter regions allowed by limits on $\delta N_{\rm eff}$
and $\sum_\nu m_\nu$ 
from the cosmic microwave background and big bang nucleosynthesis,
using a simplified yet accurate treatment of neutrino oscillations
in the early universe.  This mechanism could have an important impact
on laboratory experiments that suggest oscillations with sterile 
neutrinos.

\end{abstract}
\maketitle

{\bf Introduction.}
Short baseline (SBL) neutrino oscillation experiments at nuclear reactors
suggest at $3\sigma$ an eV-scale sterile neutrino $\nu_s$ that mixes
with $\nu_e$
\cite{Giunti:2012tn,Giunti:2012bc,Kopp:2013vaa,Dentler:2018sju,Diaz:2019fwt}.  
A persistent deficit of low-energy solar $\nu_e$ flux in gallium
experiments lends support to this interpretation. 
The NEOS \cite{Ko:2016owz} and DANSS \cite{Alekseev:2018efk}
experiments that also search for
$\nu_e$-$\nu_s$ oscillations observe features that could be 
consistent with the SBL anomalies, though are not yet conclusive.
Recent fits to the data favor a mass $m_4 = 1.1\,$eV and mixing matrix
element $U_{e4}=0.11$ \cite{Kostensalo:2019vmv}.  Moreover there are
hints from other experiments, 
LSND \cite{Aguilar:2001ty} and 
MiniBooNE \cite{Aguilar-Arevalo:2018gpe}, of $\nu_\mu\to \nu_e$
oscillations via a sterile neutrino with similar mass and mixing
parameters.  
The sterile neutrino intepretation of $\nu_\mu\to \nu_e$ is clouded
by constraints on $\nu_\mu$-$\nu_s$ oscillations from MINOS \cite{Adamson:2017uda}
and IceCube \cite{Aartsen:2017bap,Jones:2019nix}.  
In this work I
focus on the simpler $\nu_e$-$\nu_s$ scenario that could explain the SBL
deficits.  The KATRIN experiment will provide an independent probe in the
near future \cite{Esmaili:2012vg}.

A generic challenge to the existence of sterile neutrinos in 
the indicated mass and mixing range are their 
 oscillations in the early universe
that would fully equilibrate the sterile species 
\cite{Enqvist:1991qj,Dolgov:2003sg,Gariazzo:2019gyi}.
This is strongly
excluded by big bang nucleosynthesis (BBN) and cosmic microwave
background (CMB) constraints on additional effective neutrino species,
$\delta N_{\rm eff}$, as well as the sum of neutrino masses $\sum
m_\nu$.  Some means of suppressing oscillations in the early universe
while allowing them at the present time is needed.

The use of sterile neutrino interactions to inhibit
oscillations has a long history \cite{Babu:1991at,Enqvist:1992ux,
Cline:1991zb}.  With respect to the current
anomalies, refs.\ \cite{Hannestad:2013ana,Dasgupta:2013zpn} suggested that self-interactions of the
sterile neutrino could impede the oscillations and thereby satisfy the
cosmological constraints.  This mechanism is referred to as ``secret
neutrino interactions,'' despite the efforts of PRL to censor the
name.  Subsequent investigation showed that although
the self-interactions 
in this context could prevent $\nu_4$ production until freezeout of
the active neutrinos, in accordance with bounds on $N_{\rm eff}$, 
at lower temperatures their self-scattering
combines with oscillations to convert active neutrinos to $\nu_4$
and violate the CMB bound on $\sum m_\nu$.
\cite{Saviano:2014esa,Cherry:2016jol,Forastieri:2017oma,Chu:2018gxk,Song:2018zyl}. 
(An exception is found for self-interactions mediated by a
light gauge boson of mass  $\lesssim 10\,$MeV \cite{Mirizzi:2014ama}.)

It was recently pointed out that an effective realization of secret
interactions is to 
couple $\nu_s$  to ultralight bosonic dark matter $\phi$
\cite{Farzan:2019yvo}.  In that case the scalar behaves as a coherent
condensate, that has not yet started oscillating at early times.  It
can easily give a large mass to $\nu_s$ during this epoch, inhibiting
the oscillations.  Once the Hubble rate drops below $m_\phi$, the
field oscillates and redshifts with scale factor as $a^{-3/2}$ as the
universe expands.  Its contribution to $m_s$ quickly disappears,
leaving only the bare Lagrangian mass of $\sim 1$\, eV.  The ``secret
interaction'' moniker is especially appropriate in this case, since
the required coupling of $\nu_s$ to $\phi$ was shown to be exceedingly
weak, $\lambda\sim 10^{-23}$.  Similar interactions of light dark
matter to standard model neutrinos were considered with respect to
their effects on laboratory neutrino oscillations in refs.\ 
\cite{Berlin:2016woy,Krnjaic:2017zlz,Brdar:2017kbt,Liao:2018byh}.

This model is quite economical, depending only upon $m_\phi$ and the
$\nu_s$-$\phi$ coupling $\lambda$, assuming $\phi$ constitutes all of
the dark matter (DM) so that its initial amplitude is determined by
its relic density.  Equivalently, one can trade
$\lambda$ for the new contribution $m_{s,0}$ to the $\nu_s$ mass at
early times, before $\phi$ has started to oscillate.  The purpose of
this note is to determine the allowed parameter space, more
quantitatively than was done in ref.\ \cite{Farzan:2019yvo}.

{\bf Theoretical framework.}
Considering mixing between $\nu_s$ and $\nu_e$ only, the neutrino mass matrix is
\be
	\left({m_{ee}\atop m_{es}} {m_{es}\atop m_{ss}}\right)
\ee
It is assumed that $m_{ss}\gg m_{ee}$.  Then for small mixing one can 
show that $m_{es}$ is related to the mass eigenvalue $m_4\sim 1\,$eV
by
\be
	m_{es}\cong U_{e4}m_4
\ee
Fits to the SBL data favor $m_4 = 1.13\pm 0.04$,
$U_{e4}\in[0.04,\,0.13]$ \cite{Dentler:2018sju}; for definiteness
I adopt the central value $m_4 = 1.1\,$eV and $U_{e4} = 0.11$
of ref.\ \cite{Kostensalo:2019vmv}, giving $m_{es} = 0.12$\,eV and
$m_{ss} \cong m_4$.

The sterile neutrino, taken to be Majorana, 
couples to bosonic DM via
\be
	\sfrac12\lambda \bar\nu_s \phi \nu_s
\ee
leading to the effective mass $m_{\rm eff} = m_{ss} + \lambda\phi$
when DM has a VEV.  For ultralight DM, such a VEV is presumed
to exist \cite{Hu:2000ke,Hui:2016ltb}, assuming some initial value in the
early universe, that persists to account for the present relic
density.  If $\phi$ is sufficiently weakly coupled, it never 
thermalizes and remains coherent, behaving like a classical field.
Its time dependence in the expanding cosmological background is 
\footnote{The normalization is such that $\hat\phi(0) = 1$}
\be
	\phi(t) \cong 1.08\,\phi_0{\,J_{1/4}(m_\phi t)\over (m_\phi
t)^{1/4}}  \equiv \phi_0\,\hat\phi(t)
\ee
during radiation domination (when $a(t)\sim t^{1/2}$).  The relevant
combination of parameters affecting neutrino oscillations is
\be
	m_{s,0} = \lambda\phi_0
\ee
so that $m_{\rm eff} = m_{ss} + m_{s,0}\,\hat\phi(t)$.

For $t\gg m_\phi$ (but before matter-radiation equality) it can
be shown that $\rho_\phi\cong 0.37\, m_\phi^2\phi_0^2
(m_\phi t)^{-3/2}$.  Matching to the present DM density,
one finds 
\be
	\phi_0 = 1.0\times 10^{15}\,{\rm GeV}\left(10^{-15}\,{\rm eV}
\over m_\phi 	\right)^{1/4}
\ee
Such a large VEV could arise if $\phi$ is an axion-like particle,
the phase of a complex field $\Phi = |\Phi|e^{i\phi/f_\phi}$, with decay constant
$f_\phi > \phi_0$.  At early times $\rho_\phi \sim m_\phi^2\phi_0^2$ 
would be negligible compared to the energy density of radiation,
and $\phi_0$ could take random values in the interval $[0,\,2\pi
f_\phi]$.

{\bf Production of $\nu_s$.} 
Although a rigorous study of
$\nu_s$-$\nu_e$ oscillations in the early universe requires solving
the Boltzmann equation for the density matrix
\cite{Stodolsky:1986dx,Enqvist:1990ad,Sigl:1992fn}, a good approximation
can be obtained in a simpler approach, described in refs.\
\cite{Kainulainen:1990ds,Cline:1991zb}, which in some regimes leads to analytic 
results.\footnote{The quantitative agreement of the two formalisms
was recently demonstrated in ref.\ \cite{Bringmann:2018sbs}}\ \ 
The method is based upon solving the Schr\"odinger equation for the
two-state system, including an imaginary term $-i\Gamma/2$ in the Hamiltonian 
representing scattering of $\nu_e$ in the plasma, that causes
decoherence.  

The solution yields the probability for a $\nu_e$
to oscillate into $\nu_s$ between an arbitrary initial time and
a later time $t$.  From this, a rate of $\nu_s$ production is derived,
and the associated Boltzmann equation can be solved for the ratio of
$\nu_s$ occupation number relative to that of $\nu_e$, as a function of
temperature and neutrino momentum,\footnote{The factors of 2, missing
in \cite{Cline:1991zb}, account for the back-reaction from
$\nu_s\to\nu_e$ \cite{Bringmann:2018sbs}}
\be
	R \equiv {n_{\nu_s}\over n_{\nu_e}} = 
	\frac12\left(1 - \exp\left[-2\int_T^{T_i} \left( \Gamma\sin^2{\theta_m}\over
H T'\right)
	dT'\right]\right)\, .
\label{Rtpeq}
\ee
Here $\theta_m$ is the mixing angle including matter effects,
and the initial temperature $T_i$ can be taken to infinity.  The total
interaction rate, including elastic scattering, is 
\be
	\Gamma \cong \left(8 + 5\,e^{-m_e/T}\right){7\pi\over 216}\,
	G_F^2\, T^4\,p
\ee
 for a $\nu_e$ of momentum $p$ 
\cite{Cline:1991zb,Notzold:1987ik}.  The exponential factor
approximates the change at low temperatures when electrons have
decoupled from the plasma.  For relativistic neutrinos,
\be
	\sin^2 2\theta_m \cong {4 m_{es}^2\over 4 m_{es}^2 + 
	(m_{\rm eff} + 2 V_e p/m_{\rm eff})^2}
\ee
(recall that $m_{\rm eff}(t)$ is the total $\nu_s$ mass)
and
\bea
V_e &\cong& \left(2e^{-m_e/T} + \cos^2\theta_W\right)
	{14\pi\over 90\,\alpha}\sin^2\theta_W\,G_F^2 T^4 p\nn\\
 &\mp& c_1\left(2-y_n\over 2+y_n\right)\eta_b\, G_F T^3\left\{
	{1,\quad \ T\gg m_e\atop \nicefrac4{11},\ T\ll m_e}\right.
\eea
is the thermal self-energy for $\nu_e$ or $\bar\nu_e$.   The second
line incorporates effects of the electron and baryon asymmetries,
where $c_1\cong 0.95$, $\eta_b$ is the baryon-to-photon ratio,
and $y_n$ is the neutron to proton ratio as a function of 
temperature, which I take to be the standard result as shown in
ref.\ \cite{Kolb:1990vq}.  Numerically it turns out to have a negligible effect 
($<0.1\%$) on the 
following results; hence we can treat $\nu_e$ and $\bar\nu_e$ on the
same footing.

 The effective number of
extra neutrino species produced by the oscillations requires
integrating over momentum, weighted by the massless
Fermi-Dirac distribution function $f(p)$ for $\nu_e$ ,
\be
	\delta N_{\rm eff}(T) = {\int d^{\,3}p\, f(p)\, R(T,p) \over 
	\int d^{\,3}p\, f(p)}
\label{dNeff_eq}
\ee

Before numerically evaluating $\delta N_{\rm eff}$, an analytic result
can be found, in the regime where $m_\phi\lesssim 10^{-14}\,$eV,
sufficiently small that $\phi$
does not start oscillating until the integral in  eq.\ (\ref{Rtpeq})
has converged.  In that case $m_{\rm eff}\cong m_{s,0}$ can be treated as constant,
and $m_{es}^2$ can be ignored in the denominator.  The integral can be
evaluated analytically (ignoring the weak $T$-dependence of $g_*$ in the Hubble
rate $H= 1.66\sqrt{g_*}\,T^2/M_p$), to obtain
\be
	\delta N_{\rm eff} \cong \frac12\left[1-\exp\left(-{65\sqrt{7}\,\alpha^{1/2}G_F M_p m_{es}^2
	\over 576\,s_{W}(2+c_W^2)^{1/2} g_*^{1/2} m_{s,0}}\right)\right]
\ee
where $W$ denotes the Weinberg angle, $M_p$ is the unreduced Planck
mass, and $g_*\cong 10.75$ for the parameters of interest.  The
dependence on $T$ and $p$ is negligible for $T\lesssim 1\,$MeV, 
making it unnecessary to integrate over momenta.

{\bf BBN constraints.}
For larger values of $m_\phi$, the DM starts oscillating before
nucleosynthesis, which tends to activate the neutrino oscillations.
This can be compensated by also increasing $m_{s,0}$, but an
analytic treatment is no longer possible.  One should numerically
integrate over $T'$ and $p$ in eqs.\ (\ref{Rtpeq},\ref{dNeff_eq}).

Additionally for BBN, we should distinguish between oscillations that
produced a real excess in $N_{\rm eff}$, occuring before the
freezeout temperature $T_f = 3.2\,$MeV of $\nu_e$,
versus the subsequent oscillations that conserve total neutrino number
but convert some $\nu_e$ into $\nu_s$.  The reduction in $\nu_e$
density impacts BBN by changing the $p\leftrightarrow n$ equilibrium.
One can account for this by defining an effective $\delta N_{\rm
eff}^{BBN}$ \cite{Dolgov:2003sg},
\be
	\delta N_{\rm eff}^{BBN} = \frac47\left({4g_* + 7
	\delta N_{\rm eff}\over (1+Y_{\nu_e})^2} -g_*\right)
\label{effBBN}
\ee
where $g_* = 10.75$ and $Y_{\nu_e}\lesssim 1$ is the relative
abundance of $\nu_e$, reduced by oscillations between $T_f$ and
nucleosynthesis, $T_n \cong 0.1\,$MeV.  We estimate $Y_{\nu_e}$
by computing the change in $\delta N_{\rm eff}$ from the 
temperature interval $[T_f,T_n]$, using eqs.\  
(\ref{Rtpeq},\ref{dNeff_eq}).

The treatment (\ref{effBBN}) is valid when the effect of the
oscillations is to deplete the density of $\nu_e$ without 
changing its energy spectrum too dramatically.  Such spectral
distortions can change the neutron-to-proton ratio and subsequent
production of $^4$He in a way that cannot be simply modeled by a
reduction in $\nu_e$ density \cite{Kirilova:1997sv}.  

To check whether it is
justified to neglect the spectral distortion effect, I computed the collision terms of
the Boltzmann equations for $n$ and $p$ in the region of parameter
space, relevant to the BBN constraint, where $R(T,p)$ has the
strongest momentum dependence.  This occurs along the BBN exclusion
contour at its upper right-most extreme, at the lowest temperature
($T=0.1\,$MeV), where $R(T,p)|_{p=xT}\cong 0.1 \sqrt{x}$.  The thermally
averaged value is $\bar R = 0.177$.
The 
collision terms in the Boltzmann equation multiplying
the neutron and proton densities are respectively
\bea
	{\Gamma_n\over c_2} &=& \int_0^\infty dx\,{\rho f(x,y)\over 1+e^{-x-y}}
	+\int_{(1+\zeta)y}^\infty\!\!\!\!\!\!\!\!\! dx\, {(1-\rho)f(x,-y) e^{-x+y}
	\over 1 + e^{-x+y}}\nn\\
	{\Gamma_p\over c_2} &=& \int_0^\infty dx\,{(1-\rho) e^{-x-y}f(x,y)\over 1+e^{-x-y}}
	+\int_{(1+\zeta)y}^\infty\!\!\!\!\!\!\!\!\! dx\, {\rho f(x,-y) 
	\over 1 + e^{-x+y}}\nn\\
\label{nprates}
\eea
where $c_2=G_F^2(g_{\sss V}^2+3g_{\sss A}^2)T^5$, $\rho = R e^{-x}$, $f(x,y) =x^2(x+y)\sqrt{(x+y)^2+\zeta^2 y^2}$,
$y=(\Delta+m_e)/T$, $\Delta = m_n-m_p$ and $\zeta=m_e/\Delta$.
In the absence of spectral distortions, the rates are given by
$\bar\Gamma_{n,p}$ evaluated as in (\ref{nprates}) but with
$\bar \rho = \bar R e^{-x}$ in place of $\rho$.  I find that the approximation 
$\Gamma_{n,p}=\bar\Gamma_{n,p}$ is good
to $(2-3)\%$, justifying the use of eq.\ (\ref{effBBN}) for
determining the BBN constraint.

{\bf CMB constraints.}
For the CMB constraints, there is an analogous effect from late time
$\nu_e\to\nu_s$ conversions.  Even though
oscillations occuring after freezeout of $\nu_e$ should not change $\delta N_{\rm eff}$,
they can increase the sum of neutrino masses by converting some
$\nu_e$ to $\nu_s$.  Therefore the extra contribution to $\sum_\nu
m_\nu$ can be estimated as $m_{ss}$ times the asymptotic value of 
$\delta N_{\rm eff}$ that results at low $T\sim 1\,$eV, neglecting
the conservation of neutrino number below $T_f$.

\begin{figure}[t]
\begin{center}
\vskip-1.25cm
 \includegraphics[scale=0.28]{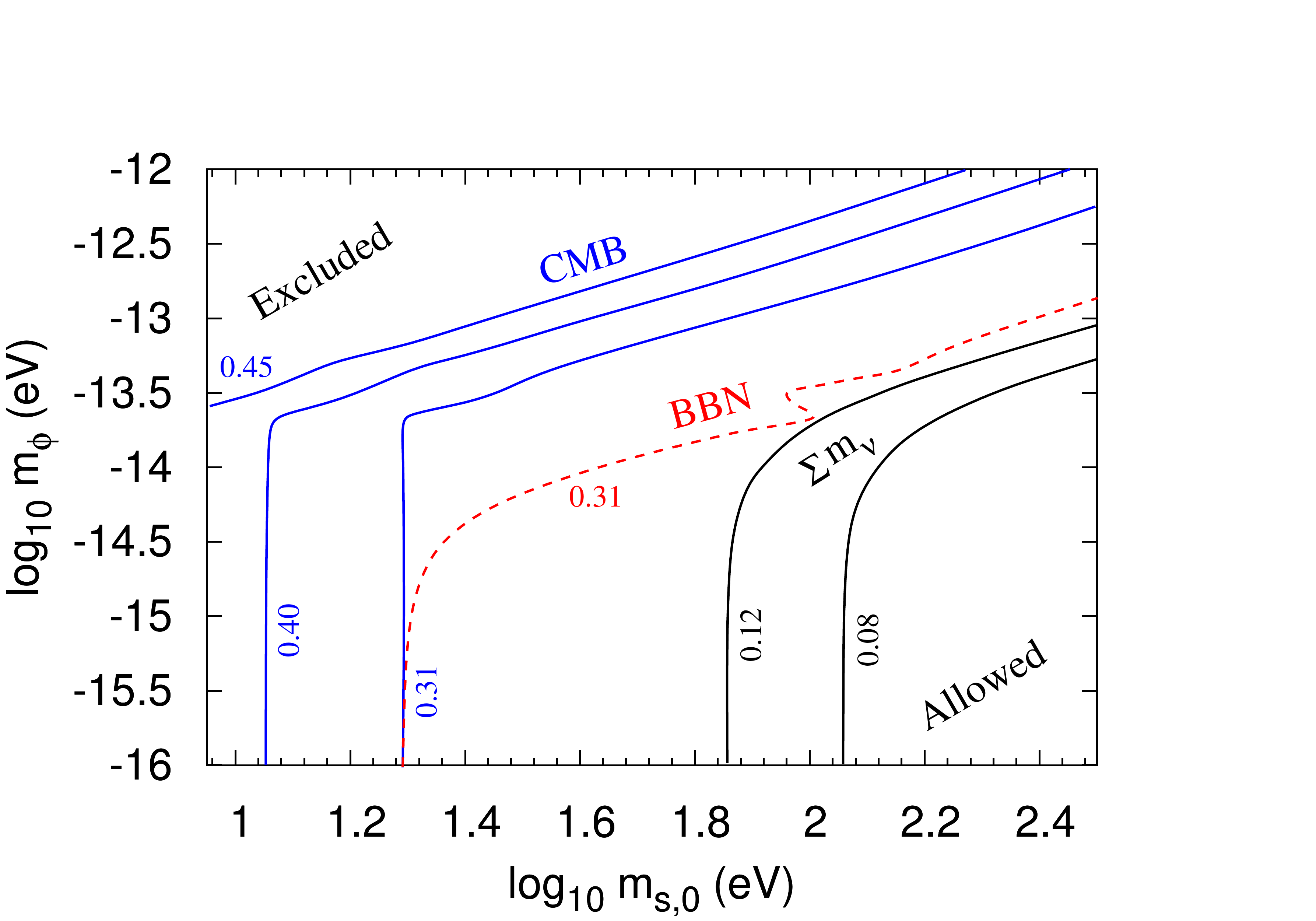}
 \caption{Contours of $\delta N_{\rm eff}$ (solid blue for CMB and
dashed red for BBN) and corresponding to 
$\sum m_\nu$ (solid black)
in the $m_{s,0}$-$m_\phi$
plane, illustrative of cosmological upper limits as described
in the text.}
 \label{fig:cont}
\end{center} 
\end{figure}

The results are shown in fig.\ \ref{fig:cont}, which displays three
contours for $\delta N_{\rm eff}$ in a region constrained by CMB measurements
\cite{Aghanim:2018eyx}.  The exact upper limit determined by the
Planck Collaboration depends upon which data sets are combined.
At 95\% c.l.\, $\delta N_{\rm eff} < 0.5$ is a typical value
(using TT+lowE or TT,TE,EE+lowE+lensing+BAO+R18), although a more 
stringent bound $\delta N_{\rm eff} < 0.23$ is derived from TT,TE,EE+lowE
alone.  To illustrate the BBN constraint I show the $2\sigma$  limit
 from ref.\ \cite{Hufnagel:2017dgo}, which is
somewhat weaker than that obtained in ref.\ \cite{Cyburt:2015mya}.
The BBN contour at $\delta N_{\rm eff} =0.31$ illustrates the effect
of conversions $\nu_e\to \nu_s$ after $\nu_e$ freezeout; for low
$m_\phi$ it coincides with the corresponding CMB $\delta N_{\rm eff}$
(since no such conversions take place), but at higher $m_\phi$,
$\delta N_{\rm eff}^{BBN}$ is seen to deviate from its CMB
counterpart, as expected.

The strongest constraint is the CMB limit on neutrino masses.  Their
sum goes as
\be
	\sum m_\nu \cong [0.06{\rm\, eV} + m_4\,\delta N_{\rm eff}] 
\ee
taking account of the standard contribution, assuming normal mass
hierarchy.  Ref.\ \cite{RoyChoudhury:2019hls} recently constrained
$\sum m_\nu < 0.145\,$eV for the normal hierarchy, 
implying $\delta N_{\rm eff}<0.08$.  This implies a lower limit on 
$m_{s,0} > 160\,$eV, hence $\lambda \gtrsim
10^{-22}\times(m_\phi/10^{-15}$\,eV)$^{1/4}$.

{\bf Discussion.}  For DM with $m_\phi \lesssim
10^{-14}\,$eV, we have seen that the cosmological analysis is 
relatively simple, since $\nu_e$ has frozen out before $\phi$ starts
to oscillate.  A favored value for $m_\phi$ from considerations of
cosmological structure formation is considerably lower, $m_\phi \sim
10^{-22}\,$eV.  In this regime, the de Broglie wavelength is so large
that structure at galactic scales can be suppressed, providing a
possible solution to the cusp/core problem of DM halos 
\cite{Hu:2000ke}.

Such light DM has an oscillation frequency of order 1\,y, which could
have interesting consequences for laboratory oscillation experiments,
if $\lambda$ is large enough to signficantly impact the effective
mass $m_{\rm eff}$ of $\nu_s$ during the timescale of the experiment.
For example, if the extra contribution to $m_{\rm eff}$ is as large
as the bare mass $m_4$, one would need $\lambda \sim 10^{-15}$,
which is technically natural since there are no significant loop
corrections.
In this situation, the usual analysis of oscillation data could lead
to ambiguous results, since the $\Delta m^2$ being fitted would be
varying in time.  This effect has already been considered with 
respect to active neutrinos coupling to $\phi$ in refs.\ 
\cite{Berlin:2016woy,Krnjaic:2017zlz,Brdar:2017kbt}.
It could be interesting to reconsider the experiments that suggest
active-sterile neutrino oscillations in this light.  A search for
time dependence of the signal has been performed by the Daya Bay
collaboration \cite{Adey:2018qsd}.

\medskip
{\bf Acknowledgment.}  I thank K.\ Kainulainen and J.\ Kopp for very useful comments on
the manuscript, and P.\ Huber for pointing out ref.\
\cite{Adey:2018qsd}.  This work was supported by NSERC (Natural Sciences
and Engineering Research Council, Canada).

\end{document}